**Thermodynamic origin of solute-enriched stacking-fault in dilute Mg-Zn-Y alloys**

M. Egami[a, b], I. Ohnuma[b, a] M. Enoki[c], H. Ohtani[c] and E. Abe[a, b]

[a] *Department of Materials Science & Engineering, University of Tokyo, Tokyo 113-8656, Japan*

[b] *Research Center for Structural Materials, National Institute for Materials Science, Tsukuba 305-0047, Japan*

[c] *Institute of Multidisciplinary Research for Advanced Materials, Tohoku University, Sendai 980-8577, Japan*

We investigate thermodynamic behaviors of dilute Mg-Zn-Y ternary alloys to form a unique solute-enriched stacking-fault (SESF), which is an intrinsic-II type stacking-fault ($I_2$-SF) enriched by the Zn and Y atoms and represents the structural-unit of the long-period stacking/order (LPSO) phase. SESF in the hexagonal-close-packed (*hcp*) Mg matrix forms a local face-centered-cubic (*fcc*) environment, and hence our thermodynamic analysis is based on the Gibbs energy comparison between *hcp* and *fcc* phases over the Mg-Zn-Y ternary composition ranges, using the calculation of phase diagrams (CALPHAD) method aided by the first principles calculations. Segregation behaviors of solute Zn/Y atoms into the SESF are firstly estimated according to the Hillert's parallel tangent law, followed by the possible disorder-order phase transformation within the SESF using the multiple-sublattice model. We find that the Zn/Y co-segregations at the SESF provide a remarkable condition that the *fcc* layers become more stable than the *hcp*-Mg matrix. Besides, within the SESF, the following spinodal-like decomposition into the Mg-rich solid-solution and the Zn/Y-rich $L1_2$-type order phase causes a significant reduction of the total Gibbs energy of the system. These thermodynamic behaviors explain fairly well a phenomenological origin of the Zn-Y clustering with the $L1_2$-type short-range order, which is known to occur for the LPSO phases and also confirmed for the present SESF by electron microscopy experiments. Therefore, strong Zn-Y interactions even in dilute conditions play a key role to stabilize firmly the SESF in the Mg-Zn-Y alloys.

## 1 Introduction

Mg alloys are the promising candidate for the next generation light-weight structural materials, and ternary Mg alloys containing a few atomic percent of transition metal (TM) and rare-earth elements (RE) have recently focused attentions because of their remarkable high-strength [1], which are provided by unique long-period structures [2–10] termed as the long-period stacking/order (LPSO) phases [6,7]. The LPSO structures are basically long-period stacking polytypes of an original hexagonal-closed-packed (*hcp*) Mg structure, for the series of which chemical order along the *c*-axis occurs so as to synchronize with the relevant stacking order [6]; synchronized LPSO polytypes of 12R [11, 12], 10H, 18R, 14H and 24R [6] have been identified so far. It is rather surprising that such largely anisotropic crystals are able to contribute for materials strengthening, since the isotropic/high-symmetry crystals with sufficient dislocation-slip systems are primary chosen for the use of structural materials. Interestingly, significant strengthening effects of the LPSO phases are realized through kink deformations [13-19], which are known to take place in favor of anisotropic crystals when the conventional dislocation motions are strongly restricted [20].

For further development of the LPSO-Mg alloys, it has become increasing importance to understand their phase formation behaviors and thermodynamic stabilities.

The LPSO structures can be well represented by a combination between the 2H-stacking (AB…) and the intrinsic-II type stacking-fault ($I_2$-SF, ABCA) [4-7] that forms local face-centered cubic (*fcc*) layers, at which the TM/RE are predominantly enriched. The TM/RE atoms are further found to be ordered within the *fcc* layers, forming the $L1_2$-type short-range order (SRO) clusters that are arranged according to an ideal superlattice dimension of $6 \times (1\bar{2}10)_{hcp}$ with respect to the fundamental *hcp* Mg lattice (6M in-plane order) [7, 21]. Given the ideal stoichiometric LPSO structure models supported with an energetic stability [7, 22], thermodynamic analyses have been attempted based on the calculation of phase diagrams (CALPHAD) method [23-25] and confirmed that the LPSO phases are indeed stable at finite temperatures.

Even though the stoichiometric LPSO structures have been well established as the sufficiently stable phase, the thermodynamic origin of the LPSO phase is not fully understood yet. It is found that the LPSO phases form with non-stoichiometry compositions widely extended into TM/RE dilute ranges, as being significantly far from the ideal LPSO stoichiometry [7, 10, 12] and bound along the definite TM/RE ratio due to the robust $L1_2$-type SRO clusters. In such dilute compositions, the LPSO phase may behave as solid solutions rather than a stoichiometric compound. Furthermore, the $I_2$-SF enriched by the TM/RE atoms, as being identical with the structural-unit of the LPSO structure, behaves like an independent phase that precipitates during aging the dilute Mg-TM-RE alloys at relatively low temperatures [26-28] (this $I_2$-SF-type precipitate is referred to as SF-γ'in Ref. 28). This fact implies that TM-RE segregation behaviors into the $I_2$-SF *fcc*-layer are the key primary events that lead to comprehensive thermodynamic understanding of the LPSO-forming ternary Mg-TM-RE system.

With above in mind, we here employ a thermodynamic analysis that considers the solute partitioning behaviors between the *hcp* and *fcc* phases in the Mg-Zn-Y alloys, based on their Gibbs energy comparison using CALPHAD over the Mg-Zn-Y ternary composition ranges. Along with such context, we here use the term "solute-enriched stacking faults; SESFs" instead of the SF-γ' previously denoted [28], in order to recollect a general view on the local *fcc*-layer formation within the *hcp* matrix. As being not limited to the solute-segregation to the SFs, we further investigate the possible disorder-order transformation within the *fcc*-SESF layers using the multiple sub-lattice *fcc* model, which is expected to unveil unique SESF stabilizations in terms of an occurrence of the $L1_2$-type SRO clusters. A validity of the calculation results will be examined by microstructure investigations of the SESFs using high-resolution electron microscopy. Finally, we will discuss on possible computation-based predictions for the new LPSO-forming systems, as being related to thermodynamic considerations of the proposed empirical rule [1].

## 2 Calculation Methods and Procedures

### *2.1 First principles calculations of the $I_2$-SF structure in hcp-Mg*

In the present thermodynamic analysis, we attempt to evaluate solute segregation behaviors at the $I_2$-SFs based on the Gibbs energy comparison between *hcp* and *fcc* phases; that is, the local $I_2$-SF configurations in the *hcp*-Mg is assumed as the *fcc*-Mg structure. Since the *c/a* ratio of the *hcp*-Mg (2H stacking), 1.623, is almost close to the ideal value 1.633, the 3R-Mg (ABC stacking) is expected to be not largely deviated from the *fcc*-Mg structure. In order to confirm the validity of this assumption, we have performed first principles calculations [29] using Vienna ab initio Simulation Package (VASP) [30, 31] for the Mg stacking polytypes, 2H, 3R and 14H with $I_2$-SF structures shown in Fig. 1, and check their local *d/a* ratio (here, *d* and *a* represent the interlayer distances and the nearest atom intervals in the close-packed layer, respectively). Table 1 summarizes the calculated *d/a* ratio, which is a half of the conventional *c*/*a* ratio for the *hcp* (2H) and provides generalized values comparable with those of 3R and 14H Mg polytypes, respectively. The *d/a* values of the 2H and 3R structures appear to be 0.8141 and 0.8147, respectively, which are close to the ideal value of 0.8165 with small differences about ~0.3 %. For the $I_2$-SF in the 14H structure, the *d/a* values deviate ranging from 0.8122 ($d_3$) to 0.8136 ($d_1$), but the differences from the ideal value are still maintained up to 0.52% ($d_3$). Therefore, it is concluded that the $I_2$-SF layers can be reasonably assumed as a local *fcc*-Mg phase.

*2.2 Cluster expansion, cluster variation method (CE-CVM) and multiple sublattice model*

For the present thermodynamic analysis, the Gibbs energies of the disordered *hcp* and *fcc* solution phases are necessary for the Mg-Zn-Y system. However, the experiment-based Gibbs energy functions are not fully available because of the experimental difficulties, such as that the Zn solubility in the *hcp*-Mg phase is too small and hence hardly measured. Besides, the *fcc*-Mg phase is hypothetic (metastable) and basically do not exist. Therefore, we employ the cluster expansion and cluster variation method (CE-CVM) [32, 33] to calculate the Gibbs energies of the disordered solution phases of arbitrary chemical compositions at finite temperatures. The free energy $F$ obtained by the CE-CVM calculation is regarded as almost equal to the Gibbs energy under the atmospheric condition and hence applied for the present thermodynamic assessment. Using an effective cluster interaction (ECI) $J_\alpha$, a cluster correlation function $\xi_\alpha$, configurational entropy term $S_\alpha$ and Kikuchi-Barker constant $\gamma_\alpha$ [33, 34] *F* is defined as follows;

$$F = \sum_\alpha J_\alpha \xi_\alpha - T \sum_\alpha \gamma_\alpha S_\alpha \quad (1)$$

In eq. (1), the summations are taken from the point cluster to the largest one, and the minimal *F* is evaluated as the variation in $\xi_\alpha$. The largest clusters chosen for the present CE-CVM calculations are the two-body cluster up to the fifteenth nearest-neighbor (about 8.6 Å in distance) and the three-body cluster up to the fifth nearest-neighbor (about 7.2 Å in distance) for the *hcp* phase the *fcc* phase, respectively. We calculate the *F* using the CE-CVM code developed by Sluiter et al. [35] and then evaluate the thermodynamic parameters used in the following Gibbs energy functions ($L^{\phi}_{i,j}$ and $L^{\phi}_{i,j,k}$ in eq. (2), details of which are described in Appendix). The Gibbs energy values by CE-CVM are then used to optimize thermodynamic parameters to formulate the Gibbs energy as a function of the composition and temperature. The thermodynamic parameters are calculated using ThermoCalc ver.S [36], and PANDAT 2017 is employed for the spinodal line calculation. The Gibbs energy of the disordered solution phases (liquid, *hcp* and *fcc*) is described by the sub-regular solution approximation using following Redlich-Kister-Muggianue polynomial formulae for a $\phi$ phase;

$$G^{\phi} = \sum_i x_i {}^{\circ}G^{\phi}_i + RT \sum_i x_i \ln x_i + \sum_{i>j} x_i x_j L^{\phi}_{i,j} + x_i x_j x_k L^{\phi}_{i,j,k} \quad (2),$$

where $R$, $T$, $x_i$ and $°G_{\mathrm{i}}^{\phi}$ represent the gas constant, absolute temperature, mole fraction of element $i$ and the lattice stability parameter taken from the SGTE database [37], respectively.

In order to investigate further the thermodynamic origin of the Zn-Y SRO clusters in the *fcc*-SESF layers, we investigate possible disorder-order transformations based on the Gibbs energy calculations of the representative ordered *fcc* structures, $L1_2$-$A_3B$, $L1_0$-AB, $L1_2$-$AB_3$ (Fig. 2), using a single formula by the split compound energy formalism (s-CEF) with four sub-lattice model, as following equation;

$$G_{\mathrm{m}}^{\mathrm{order}} = G_{\mathrm{m}}^{\mathrm{fcc}}(x_i) + \Delta G_{\mathrm{m}}^{\mathrm{order}}\left(y_i^{(s)}\right)$$

$$= G_{\mathrm{m}}^{\mathrm{fcc}}(x_i) + G_{\mathrm{m}}^{\mathrm{4SL}}\left(y_i^{(s)}\right) - G_{\mathrm{m}}^{\mathrm{4SL}}\left(y_i^{(s)} = x_i\right) \quad (3)$$

where, $G_{\mathrm{m}}^{\mathrm{fcc}}(x_i)$, $y_i^{(s)}$ and $\Delta G_{\mathrm{m}}^{\mathrm{order}}(y_i^{(s)})$ represent the Gibbs energy of the *fcc* disordered solution phase described by eq. (2), the site fraction of element $i$ in the $s$-th sublattice and the contribution of the ordered *fcc* phases described by four sub-lattice model, respectively [38]. The Gibbs energy using the four sub-lattice model is described as;

$$G^{\mathrm{4SL}}(y_i^{(s)}) = \sum_i \sum_j \sum_k \sum_l y_i^{(1)} y_j^{(2)} y_k^{(3)} y_l^{(4)}\, G_{i,j,k,l}^{\mathrm{4SL}} + 0.25RT \sum_{t=1}^{4} \sum_i y_i^{(t)} \ln y_i^{(t)}$$

$$+ \sum_{t=1}^{4} \sum_{u>t} y_A^{(\mathrm{t})} y_B^{(\mathrm{t})} y_A^{(\mathrm{u})} y_B^{(\mathrm{u})} \left( \sum_{v \neq t,u} \sum_{w \neq t,u,v} \sum_{p=A}^{B} \sum_{q=A}^{B} y_p^{(\mathrm{v})} y_q^{(\mathrm{w})} L_{A,B:A,B:p:q}^{\mathrm{4SL}} \right) \quad (4)$$

where (i) the first term represents a mechanical mixture of all the stoichiometric compounds; (ii) the second term corresponds to the mixing entropy summed up for all the sub-lattices; (iii) the third term expresses the excess interaction energies that consider reciprocal mixing in the two of four sub-lattices simultaneously. $G_{\mathrm{m}}^{\mathrm{4SL}}\left(y_i^{(s)} = x_i\right)$ in eq. (3) corresponds to the Gibbs energy calculated by eq. (4) when all the $y_i^{(s)}$ values equal $x_i$, *i.e.*, the Gibbs energy of the fictitious disordered *fcc* structure. The thermodynamic parameters, $G_{i,j,k,l}^{\mathrm{4SL}}$ and $L_{A,B:A,B:p:q}^{\mathrm{4SL}}$ in eq. (4), are evaluated using the formation enthalpy of the $L1_2$-$A_3B$, $L1_0$-AB and $L1_2$-$AB_3$ compounds calculated by VASP, details of which are described in Appendix.

### *2.3 Evaluation of solute segregation by extended Hillert's parallel tangent law*

Given the Gibbs energies of the *hcp* and *fcc* disordered solution phases as the functions of composition and temperature, the thermodynamic origin of the SESF can be investigated according to the Hillert's parallel tangent law [39]. This is schematically explained with the case of solute segregations at grain boundaries (GBs), as shown in Fig. 3. In this scheme, random atomic configurations at the GBs are assumed as being analogous to those in a liquid phase, and consider the Gibbs energies of the matrix and liquid phases. Composition at the GBs, $x^{GB}$, can be estimated by the tangential line that is drawn parallel to the one defined at the composition of the matrix, $x^{matrix}$, as shown in Fig. 3, in the sense that the chemical potential differences to be equal for each of the elements those in the GB and matrix. Therefore, the following equation should be satisfied;

$$\mu_{\mathrm{A}}^{\mathrm{GB}} - \mu_{\mathrm{A}}^{\mathrm{matrix}} = \mu_{\mathrm{B}}^{\mathrm{GB}} - \mu_{\mathrm{B}}^{\mathrm{matrix}} \tag{5}$$

where $\mu_i^{\theta}$ represents the chemical potential of an element $i$ in the θ phase. The solute segregations at the $I_2$-SFs in the *hcp*-Mg matrix can be evaluated along the same manner, by assuming that the $I_2$-SF environments are equivalent to the *fcc* structure as described in the previous section. For the present Mg-Zn-Y ternary system, the parallel tangent lines are extended to the parallel tangent planes drawn on the *G* surfaces of the *hcp-fcc* phases, under the thermodynamic conditions defined by the following equations.

$$\mu_{\mathrm{Mg}}^{\mathrm{fcc}} - \mu_{\mathrm{Mg}}^{\mathrm{hcp}} = \mu_{\mathrm{Zn}}^{\mathrm{fcc}} - \mu_{\mathrm{Zn}}^{\mathrm{hcp}} = \mu_{\mathrm{Y}}^{\mathrm{fcc}} - \mu_{\mathrm{Y}}^{\mathrm{hcp}} \tag{6}$$

With a certain *fcc* volume fraction, the mass conservation described by the following equation is taken into account.

$$M_{\mathrm{f}}^{\mathrm{fcc}}\left(x_{\mathrm{A}}^{0} - x_{\mathrm{A}}^{\mathrm{hcp}}\right) = M_{\mathrm{f}}^{\mathrm{hcp}}\left(x_{\mathrm{A}}^{\mathrm{fcc}} - x_{\mathrm{A}}^{0}\right) \qquad (7),$$

where $M_{\mathrm{f}}^{\mathrm{fcc}}$ and $M_{\mathrm{f}}^{\mathrm{hcp}}$ represent the molar fraction of the *fcc*-SESF layers and the *hcp*-matrix respectively, and A represents either Zn or Y in the present case. $x_{\mathrm{A}}^{0}$ denotes the mole fraction of the solute element A in the initial *hcp*-Mg phase, and $x_{\mathrm{A}}^{hcp}$ and $x_{\mathrm{A}}^{fcc}$ represent respectively that in the *hcp*-Mg matrix and the *fcc*-SESF after the solute segregations. Under the constraint condition of eq. (7), the partitioned Zn/Y compositions at the *fcc*-SESF layers can be determined by a numerical calculation based on the parallel tangent law described by eq. (6).

## 3 Experimental Procedures

In the present work, we investigate the microstructural details of the SESFs using high-resolution electron microscopy and compare with the thermodynamic modeling results. A master alloy ingot with a nominal composition Mg-1at.% Zn-2at.% Y (hereafter the composition is denoted as $Mg_{97}Zn_1Y_2$ in at.%) was prepared by high-frequency induction melting of pure Mg, Zn and Y metals in a carbon crucible. A piece of the master ingot was sealed in a Pyrex tube filled with an argon atmosphere after evacuation to pressures lower than $3\times10^{-3}$ Pa, and then solution-treated at 823K for 8 hours followed by quenching into water. After the solution-treatment, the alloy pieces were annealed at 523K for 1 week to promote the SESF growth, followed by quenching into water. Compositions of α-Mg grains were measured using scanning electron microscopy (SEM, Hitachi S-4200 FE-SEM) equipped with energy dispersive X-ray spectroscopy (EDS) at the accelerating voltage of 15 kV. Thin-foil specimens for scanning transmission electron microscope (STEM) observations were prepared by mechanical polishing and standard argon ion milling technique. High-angle annular dark-field (HAADF) STEM observations were performed by an aberration-corrected 200 kV microscope (JEOL JEM-ARM200F) equipped with a cold field-emission gun. Annular detectors were set to collect electrons scattered at angles between 90 and 370 mrad for high angle annular dark field (HAADF) imaging. For HAADF imaging the annular detector was set to collect electrons scattered at angles higher than 90 mrad, which is sufficiently high to obtain the atomic-number dependent Z-contrast.

## 4 Results

### *4.1 Microstructure characterizations of the SESFs*

Figure 4a shows the SEM image of the present $Mg_{97}Zn_1Y_2$ alloy aged at 523K for a week. The microstructure consists of α-Mg grains (50 ~ 100 μm), which are surrounded by the retained LPSO phase at grain boundaries [3]. That is, these LPSO phases primarily formed during solidifications and did not fully dissolve into the α-Mg matrix even during the solution treatment (823K for 8 hours). While, the α-Mg phase forms a supersaturated solid-solution by the solution treatments, and the SESFs precipitations have been promoted during aging at low temperature (523K) to form a fine lamellar structure, as clearly revealed by the corresponding HAADF image in Fig. 4b. The bright lines are indeed confirmed as $I_2$-SFs accompanying the Zn/Y enrichment; i.e., the SESF shown in Fig. 4c. Further details of the Zn/Y configurations within the SESF will be described later.

Average composition of the α-Mg grains including the SESFs is determined to be Mg-0.6±0.2at.%Zn-1.2±0.3at.%Y by SEM-EDS analysis. In the present thermodynamic modeling, the total Zn/Y contents in the α-Mg grains are assumed to be constant before and after the aging; i.e., all the SESFs have grown from the supersaturated α-Mg during aging without any Zn/Y supplies from the LPSO phase at grain boundaries (though, in real alloys, a certain amount of Zn/Y may have been supplied from the LPSO phases). Therefore, the present analysis treats the α-Mg - SESF equilibrium system with the total composition Mg-0.6at.%Zn-1.2at.%Y at 523K. Given the fact that each of the bright-line contains two *fcc* layers, volume fraction of the local *fcc* layers ($V_{\mathrm{f}}^{fcc}$) in the α-Mg matrix can be evaluated from the HAADF images. The average distance between the bright lines is estimated to be approximately 17.0 nm from a number of the HAADF images (total surveyed area is about $\sim 1.0 \times 10^{6}$ nm$^2$), providing $V_{\mathrm{f}}^{fcc}$ to be approximately 3.0% (i.e., 0.52nm/17nm) by assuming that all the observed *fcc* layers penetrate through the specimen. As a counterpart, the volume fraction of the *hcp* matrix, $V_{\mathrm{f}}^{hcp}$, is given as 97.0%. These volume fractions will be used in the following thermodynamic calculations.

*4.2 Tuning the thermodynamic parameters*

We firstly attempt to tune the thermodynamic parameters for the Mg-Y, Mg-Zn and Y-Zn binary systems, and the results are shown in Figs. 5 a-c, d-f and g-i, respectively. The determined thermodynamic parameters are summarized in Table 2. During the assessments, the Gibbs energies of *hcp* and *fcc* solid-solution phases for each of the binary systems have been firstly evaluated based on CE-CVM, as shown by the dot-plot in Figs. 5 (c, f, i). These CE-CVM-based Gibbs energies are further modified by conventional linear transformations (i.e., only the composition-dependent terms in eq. (2) are transformed), such that the calculated phase diagrams (Figs. 5a, d, g) reproduce well the experimental stable diagrams. The thermodynamic parameters for the other phases, liquid and intermetallic compounds, are referred from the previous work [40]. Formation enthalpies of the several ordered *fcc* phases are derived from first principles calculations and listed in Table 3 for each of the binary and the ternary systems. During their optimizations, the lattice parameters of the $L1_0$-type structure are fixed as a cubic to avoid distortions to a tetragonal lattice. According to eqs. (A. 8') - (A. 10') in the Appendix, nine thermodynamic parameters of the ordered *fcc* phases are determined, as summarized in Table 4. Consequently, the *fcc*-based metastable phase diagrams for each of the binary systems have been successfully calculated, as shown in Figs. 5 (b, e, h). It is worth noting that, among the ordered *fcc*-phases, the $L1_2$-$Zn_3Y$ phase is emerged as the highest stable one, as maintained up to 1600K (Fig. 5h).

Gibbs energies of *hcp* and *fcc* solid-solution phases in the Mg-Zn-Y ternary system results are shown in Figs. 6 (a-c). The CE-CVM-based Gibbs energies have been also modified by conventional linear transformations along with the same manner used for the binary systems. Accordingly, the calculated Mg-Zn-Y stable phase diagram at 523K (Fig. 6d) reproduces fairly well the experimental ternary diagram. It is noteworthy that no *fcc*-based phases remain to be stable, even though they are indeed involved in the present calculations.

### *4.3 Solute atom behaviors into/within the SESFs*

Given all the necessary thermodynamic parameters described above, Gibbs energy surfaces of the *hcp* and *fcc* solid-solution phases are successfully derived as shown in Fig. 7a. Overall, both the *hcp* and *fcc* surfaces reveal convex downward from the each corner and reach the minimum somewhere at the intermediate Y-Zn line.

Interestingly, at the particular composition ranges the surfaces reveal convex slightly upward, as indicated by the bold arrow in Fig. 7a, around where the *fcc* surface becomes lower than that of the *hcp*. Details of this feature are described in the followings. According to the extended Hillert's parallel tangent law under the mass conservation condition of $V_\mathrm{f}^{hcp}:V_\mathrm{f}^{fcc}$ = 0.97 : 0.03 and the total composition $Mg_{98.2}Zn_{0.6}Y_{1.2}$, compositions of the α-Mg matrix ($x_{hcp}$) and the SESF ($x_{fcc}$) are successfully derived to be $Mg_{99.0}Zn_{0.05}Y_{1.0}$ and $Mg_{72.2}Zn_{18.5}Y_{9.3}$, respectively. The corresponding tangent planes and points are drawn in Fig. 7a, and its vertical cross section to involve the both tangent points is shown in Fig. 7b, crossing with the edge compositions represented as A ($Mg_{99.0}Y_{1.0}$) and B ($Mg_{0.1}Zn_{70.5}Y_{29.4}$). Note that the $G^{fcc}$ curve indeed becomes lower than the $G^{hcp}$ curve at the composition range shown by grey in Fig. 7b; this is seen more obviously in Fig. 7c, where the *G* curves are redrawn by taking the base-line (dot-line in Fig. 7b) to be flat, which assumes the $G^{hcp}$ to be equivalent at the both sides of A and B. The corresponding tangent points and compositions of the *hcp*-matrix and the *fcc*-SESF layers, $x_{hcp}$ and $x_{fcc}$, are enlarged and shown in Figs. 7 d and e, respectively. It is emphasized again that the *fcc* phase becomes stable against the *hcp* phase ($G^{hcp} > G^{fcc}$) around the *fcc*-SESF composition range, despite the fact that the *fcc* configuration originally represents the faulted region in the *hcp*-Mg matrix. Therefore, as being not like typical defects described in Fig. 3, the driving force can be possibly given to promote precipitate and grow of the *fcc* layers aided by solute-segregations, as indicated in Fig. 7c. It is also remembered that both the $G^{hcp}$ and $G^{fcc}$ curves reveal convex upwards around the SESF compositions, suggesting a further phase-separation tendency.

Now we describe the solute Zn/Y behaviors within the *fcc*-SESF layers. Figure 8a shows the isothermal section of the metastable *fcc* solid-solution Mg-Zn-Y phase diagram at 523K. Note that a significant miscibility gap island appears to separate between the Mg-rich and Zn/Y-rich phases, as being caused by a largely negative *G* curves for the Zn-Y system (Fig. 5i) compared with those of the Mg-Y and Mg-Zn systems [41]. As a result, the spinodal decomposition line [42, 43] can be traced as indicated by arrow in Fig. 8a, and the determined SESF composition ($x_{fcc}$) is found to be indeed located within the spinodal range. Therefore, the solute atoms in the *fcc*-SESF layers spontaneously separate into the two phases along with the $G^{fcc}$ minimum, and their compositions are found to be $Mg_{98.7}Zn_{1.3}$ ($x_{fcc}^{\mathrm{Mg-rich}}$) and $Mg_{1.4}Zn_{64.4}Y_{34.2}$ ($x_{fcc}^{\mathrm{Zn/Y-rich}}$), respectively. This is again seen more clearly in the cross-section $G^{fcc}$ curve (Fig. 8b) redrawn in Fig. 8c, connecting tie-line between the separated compositions and crossing with the edge compositions represented as A' ($Mg_{99.1}Y_{0.9}$) and B' ($Zn_{68.1}Y_{31.9}$).

Here, we further consider the disorder-order effect in the *fcc*-SESF phase. Figure 8d shows the isothermal section of the metastable *fcc* phase diagram including the *fcc*-based order phases. As seen for the corresponding cross-sectional *G* curves both for the solid-solution (disordered) and the $L1_2$ ordered *fcc* phases in Fig. 8e, the system indeed gains a further stability by the $L1_2$ order process; again this is seen more clearly for the *G* curves redrawn in Fig. 8f. As a consequence along with the $G^{L1_2}$ minimum, the most stable conditions of the *fcc*-SESF are given with the compositions $Mg_{95.7}Zn_{2.8}Y_{1.5}$ ($x_{fcc}^{\text{Mg-rich}}$), $Mg_{1.0}Zn_{64.7}Y_{34.3}$ ($x_{L1_2}^{\text{Zn/Y-rich}}$) and the edge compositions A" ($Mg_{99.9}Y_{0.1}$) and B" ($Zn_{66.6}Y_{33.4}$).

## 5 Discussion

### *5.1 Thermodynamic stabilizations of the SESFs in the Mg-Zn-Y system*

On the basis of the present thermodynamic calculations described above, the *fcc*-SESF stabilization in the *hcp*-Mg matrix can be phenomenologically summarized as schematic illustrations in Fig. 9, which involves two prominent phase separation processes. When the stacking faults ($I_2$-SFs) have been somehow introduced (Fig. 9a), solute Zn/Y atoms tend to segregate into the $I_2$-SF layers to form the SESF (Fig. 9b). It should be emphasized that, with the determined segregate-composition $x_{fcc}$ (Fig. 7c), the present SESF formation can be spontaneously driven by the negative $\Delta G$ ($G^{hcp} > G^{fcc}$), which is not like the general segregation behaviors to the faults (i.e., $G^{faults}$ is generally located higher than $G^{matrix}$, such as described in Fig. 3). Followed by the SESF formation, the spinodal decomposition into the Mg-rich and Zn/Y-rich phases (Figs. 8a-c) takes place within the *fcc*-SESF layers (Fig. 9c), which then further promote the disorder-order transformation for the Zn/Y-rich regions when its composition shifts into the range $G^{fcc} > G^{L1_2}$ (Fig. 8f). After all of these continuous events, the *fcc*-SESF layers are significantly stabilized. In brief sum;

i) Solute segregations into the $I_2$-SFs (local *fcc* layers) to form the SESFs,

ii) Spinodal decomposition into the Mg-rich and Zn/Y-rich phases within the SESFs,

iii) Disorder – $L1_2$-order transformation for the Zn/Y-rich regions within the SESFs.

Because of these specific thermodynamic conditions for the Mg-Zn-Y system, the SESFs are able to precipitate and grow during aging under the occasional local conditions either by the solute-enriched *hcp* or the pre-existing stacking fault regions. The LPSO/$\alpha$-Mg interface may be a likely candidate for such regions, initiating either the solute-enrichment or fault generations into the host *hcp*-Mg structure. Once the nucleation of SESFs has been somehow realized, their growth can be spontaneously driven by the negative $\Delta G$. During the growth, the solute can be supplied from the $\alpha$-Mg matrix at the growth edge, where the local stacking change takes place simultaneously according to the so-called ledge-mechanism [44-46], as known for a lateral/thickening growth of platelet precipitations.

*5.2 Zn-Y order within the SESFs; the $L1_2$-type short-range order clusters*

As described earlier in the introduction, the present SESFs are the structural-unit of the LPSO phase, and therefore the details of a Zn-Y order can be discussed in the analogue of those in the LPSO structures. The characteristic Zn-Y configurations that have been emerged by the LPSO analysis are the $L1_2$-type short-range order (SRO) $Zn_6Y_8$ clusters; note that the $L1_2$-SRO clusters occur as isolated entity in the *hcp* superlattice ($2\sqrt{3}a \times 2\sqrt{3}a$) [7], and hence the Zn/Y ratio appears to be 3:4 ($Zn_6Y_8$) as deviated from the original $L1_2$-order of 3:1 ($L1_2$-$Zn_3Y$). In fact, it is experimentally confirmed that the LPSO composition range shows up with a definite pseudo-binary behavior well bound along the Zn/Y ratio of 3/4 [7, 10, 12], supporting well the robust $Zn_6Y_8$-cluster characteristics [47]. Although the present calculations cannot reproduce precisely the experimental Zn/Y ratio (~3/4) due to a limitation of the principle framework (i.e., we considered only the simple *fcc* order and did not the superlattice $Zn_6Y_8$-cluster order), the pseudo-binary behaviors are nevertheless emerged with the definite tie-line between the almost pure-Mg and the $L1_2$-order phases (Figs. 8d-f). Again, this explains well the phenomenological thermodynamic insights of the present Mg-Zn-Y system.

We now describe the experimental investigations of the cluster characteristics in the present *fcc*-SESFs. Fig. 10a shows atomic-resolution HAADF-STEM image of the SESF taken along the $[10\bar{1}0]$ direction, from which the local $L1_2$-clusters can be traced by highlighting the Zn/Y positions [7, 9, 10]. At some places, the bright-dot arrangements representing $L1_2$-$Zn_6Y_8$ clusters (Fig. 10d) are clearly seen as exemplified by rectangles, confirming that there indeed form the Zn-Y clusters in the SESF. In the corresponding fast Fourier transformation (FFT) pattern in Fig. 10b, a diffuse peak relevant to the $L1_2$-type SRO cluster appears across the $(11\bar{2}0)$ direction, whose intensity profile is shown at the bottom. Interestingly, the diffuse peak appears to be not consistent with the simple double-ordered *hcp* of $d(11\bar{2}0)$ = 3.2Å but at a significantly deviated position approximately 2.8Å. The inverse Fourier image using the diffuse peak of 2.8Å correlation is shown in Fig. 10c, where the distinct fringe contrasts have been well reproduced at the original $L1_2$-$Zn_6Y_8$ cluster positions. Since the $L1_2$-$Zn_6Y_8$ configurations are shown to be significantly relaxed from the original *fcc* positions after energetic tuning of the LPSO structure [7], the correlation-length of 2.8Å can be attributed reasonably to the Y-Y distances after the cluster relaxation, as shown in Fig. 10d. On these bases, almost all the cluster-characteristics in the LPSO phase are also realized in the present SESFs. Here, it should be mentioned that the relaxation of the $L1_2$-$Zn_6Y_8$ cluster provides significantly large stability; it turned out to be -66.7 kJ/mol gain for the 18R-LPSO phase (four $Zn_6Y_8$-clusters per unit-cell of 18R-$Mg_{116}Zn_{12}Y_{16}$ [7]). Therefore, in addition to the thermodynamic stabilization processes described as i) - iii) in the previous chapter, the SESFs in the real alloys are further stabilized by the cluster relaxation process.

*5.3 Possible predictions of the LPSO/SESF forming Mg-$X_1$-$X_2$ ternary alloys*

Finally we briefly discuss on how to search the Mg-$X_1$-$X_2$ ternary alloys that possibly form the LPSO phase as well as unique SESFs. Empirically, the thermodynamic criteria were suggested for the $X_1$ and $X_2$ elements to satisfy [1]; i) they have certain solubility to the *hcp*-Mg matrix, and ii) their mixing enthalpy ($\Delta H_{mix}^{X_1-X_2}$) is largely negative. The present work has clarified detailed thermodynamic conditions in relation to these key factors. First, the solute $X_1$-$X_2$ co-segregation is able to provide the reversed matrix-fault Gibbs energy ranges; i.e., $I_2$-SFs can be more stable than the *hcp*-matrix ($G^{hcp} > G^{fcc}$ in Fig. 7c) at the segregation Zn/Y composition.

Second, the segregated $X_1$ and $X_2$ elements at the fault cause additional energy-gain events; i.e., within the present SESFs, spinodal decomposition and the following disorder-order transformation takes place to reduce significantly the Gibbs energy. We note that these step-by-step events (Fig. 9) would lead to spontaneous formation of highly solute-enriched regions in the faults, typically forming the robust solute clusters driven by large $\Delta H_{mix}^{X_1-X_2}$ and firmly stabilize the faults. With these in mind, it is possible to predict the $X_1$-$X_2$ element combinations that show similar Gibbs energy behaviors with the present Zn-Y elements, according to the CALPHAD aided by first principles calculations as demonstrated in the present work. Conversely, it may also provide important insights to understand the thermodynamic details of the Mg-$X_1$-$X_2$ alloys that are critical to form either LPSO or similar long-period structure phases [48-50].

These thermodynamic criteria may further be extended for the other metal matrix for exploring the novel LPSO-related and SESF-layer structures; e.g., for *hcp*-Ti, *fcc*-Al alloys. Even though the stacking-fault energy is known to be considerably high for the Al alloys, some of the solute combinations might be able to stabilize the stacking faults of local *hcp* regions, as being in the similar co-segregation manner described for the present Mg-Zn-Y ternary alloys. To search such candidate elements, an empirical machine learning as well as first principles computations will play a key role.

Controlling layer-arrangements of the highly stable SESFs will lead to a novel concept of microstructure design, covering from the LPSO structure to a general layer-structure composed of randomly/sparsely distributed SESFs. Suppose that the kink deformation can be effectively induced by such sparsely-distributed SESFs, the Mg alloys can be more dilute and hence lighter by keeping their strength comparable to that of the LPSO-type alloys. We now attempt to develop novel high-strength Mg alloys along this concept, and the results will soon be described elsewhere.

## 6 Summary

In the present study, we have investigated the thermodynamic origins of the SESFs formed in the dilute Mg-Zn-Y alloys, based on the Gibbs energy comparison between *hcp* and *fcc* phases over the ternary composition ranges, using the CALPHAD method aided by the first principles calculations. The major results are summarized as follows.

1. First-principles calculations show that there are no significant fundamental lattice distortions between the 2H (*hcp*), 3R (*fcc*) and 14H stacking ($I_2$-SF) Mg structures.

Therefore, the local $I_2$-SF configurations in the *hcp*-Mg can reasonably be assumed as the *fcc*-Mg phase. On this basis, we have evaluated solute segregation behaviors at the $I_2$-SFs based on the Gibbs energy comparison between *hcp* and *fcc* phases.

2. Thermodynamic parameters of the *hcp* and *fcc* disordered solution phases in the Mg-Zn-Y system have been evaluated using the CE-CVM, and the representative ordered *fcc* structures are incorporated for the present Gibbs energy calculations using a single formula with four sub-lattice model. On the basis of SEM/STEM observations, the entire composition of the dilute Mg-Zn-Y grain and the volume fraction of the SESFs have been experimentally determined.
3. Given all the necessary parameters, the Gibbs energy surfaces of the *hcp* and *fcc* phases have been successfully derived over the ternary composition range. According to the extended Hillert's parallel tangent law, the solute Zn/Y compositions at the *fcc*-SESF layers in the *hcp*-Mg matrix are evaluated to be approximately $Mg_{72.2}Zn_{18.5}Y_{9.3}$. Interestingly, at this composition the *fcc*-SESF layers are found to become more stable than the *hcp*-Mg matrix ($G^{hcp} > G^{fcc}$).
4. Within the *fcc*-SESF layers, the spinodal decomposition into the Mg-rich and Zn/Y-rich phases takes place, the latter of which then further undergoes disorder-order transformation to form the $L1_2$-type Zn-Y phase. Atomic-resolution

HAADF-STEM observations of the SESFs have confirmed that the Zn/Y atoms form the $L1_2$-type SRO-$Zn_6Y_8$ clusters, which are found to be significantly relaxed and deviated from the original *fcc* configurations. After all of these events, the *fcc*-SESF layers are significantly stabilized.

5. The present work has clarified key thermodynamic conditions to serach the Mg-$X_1$-$X_2$ ternary alloys that possibly form the LPSO phase as well as unique SESFs. First, the solute $X_1$-$X_2$ co-segregation is able to provide the reversed matrix-fault Gibbs energy ranges. Second, the segregated $X_1$ and $X_2$ elements at the fault cause additional energy-gain events, such as phase separation and/or disorder-order transformation. These thermodynamic criteria may further be extended for the other metal matrix for exploring the novel LPSO-related and SESF-layer structures; e.g., for *hcp*-Ti, *fcc*-Al alloys.

**Acknowledgment**

This study is supported by JSPS KAKENHI for Scientific Research on Innovative Areas "Materials Science of a Mille-feuille Structure (Grant Numbers JP18H05475, JP18H05476, JP18H05479)", and "Nanotechnology Platform" of the MEXT, Japan.

## Appendix

A.1. Gibbs energy of *hcp* and *fcc* solution phases

$L^{\phi}_{i,j}$ and $L^{\phi}_{i,j,k}$ in eq. (5) represent the interaction parameters in the *i-j* binary and *i-j-k* ternary systems, respectively, which are described as following equations;

$$L^{\phi}_{i,j} = \sum_{n} (x_i - x_j)^n \, {}^{n}L^{\phi}_{i,j} \qquad \text{(A.1)},$$

$$L^{\phi}_{i,j,k} = x_i \, {}^{0}L^{\phi}_{i,j,k} + x_j \, {}^{1}L^{\phi}_{i,j,k} + x_k \, {}^{2}L^{\phi}_{i,j,k} \qquad \text{(A.2)},$$

where each $L$ parameter in eqs. (6) and (7) depends on the absolute temperature, $T$, as

$$L = a + bT \qquad \text{(A.3)}$$

On the other hand, the parameters for the ordered structures are written as eq. (7). Assuming that the reciprocal parameter value does not depend on the elements occupying other two sub-lattices, following relationships among the reciprocal parameters are fulfilled:

$$L^{4\mathrm{SL}}_{A,B:A,B:A:A} = L^{4\mathrm{SL}}_{A,B:A,B:A:B} = L^{4\mathrm{SL}}_{A,B:A,B:B:A} = L^{4\mathrm{SL}}_{A,B:A,B:B:B} = L^{4\mathrm{SL}}_{A,B:A,B:*:*} \qquad \text{(A.4)},$$

where the asterisk "*" represents all the possible occupations of either A or B element. Due to the crystallographic symmetry, the value of the equivalent parameters should not differ from each other depending on the permutation of the occupation among four sub-lattices, *i.e.*, examples are;

$$G^{4\mathrm{SL}}_{A:A:A:B} = G^{4\mathrm{SL}}_{A:A:B:A} = G^{4\mathrm{SL}}_{A:B:A:A} = G^{4\mathrm{SL}}_{B:A:A:A} = G^{4\mathrm{SL}}_{A_3B} \qquad \text{(A.5)}$$

$$G^{\mathrm{4SL}}_{A:A:B:B} = G^{\mathrm{4SL}}_{A:B:A:B} = G^{\mathrm{4SL}}_{A:B:B:A} = G^{\mathrm{4SL}}_{B:A:A:B} = G^{\mathrm{4SL}}_{B:A:B:A} = G^{\mathrm{4SL}}_{B:B:A:A} = G^{\mathrm{4SL}}_{A_2B_2} \quad (\mathrm{A.6})$$

$$L^{\mathrm{4SL}}_{A,B:A,B:*:*} = L^{\mathrm{4SL}}_{A,B:*:A,B:*} = L^{\mathrm{4SL}}_{A,B:*:*:A,B} = L^{\mathrm{4SL}}_{*:A,B:A,B:*} = L^{\mathrm{4SL}}_{*:A,B:*:A,B} = L^{\mathrm{4SL}}_{*:*:A,B:A,B} \quad (\mathrm{A.7})$$

The Gibbs energies of stoichiometric compounds can be described using the nearest-neighbor bond energies. The $A_3B$ and $AB_3$ compounds consist of the same three atoms and another one occupied the four sub-lattices, *i.e.*, they form three A-B bonds between unlike atoms. In the case of the $A_2B_2$, four A-B bonds exist. Thus, using a single bond energy, $w_{A:B}$, the Gibbs energies of the binary stoichiometric compounds can be expressed as;

$$G^{\mathrm{4SL}}_{A_3B} = 3w_{A:B} \quad (\mathrm{A.8})$$

$$G^{\mathrm{4SL}}_{A_2B_2} = 4w_{A:B} \quad (\mathrm{A.9})$$

$$G^{\mathrm{4SL}}_{AB_3} = 3w_{A:B} \quad (\mathrm{A.10})$$

The bond energy, however, may depend on their environment which has either more A atoms than B or the opposite. Therefore, correction factors are added [27] as;

$$G^{\mathrm{4SL}}_{A_3B} = 3w_{A:B} + 3\alpha_{A_3B} \quad (\mathrm{A.8'})$$

$$G^{\mathrm{4SL}}_{A_2B_2} = 4w_{A:B} \quad (\mathrm{A.9'})$$

$$G^{\mathrm{4SL}}_{AB_3} = 3w_{A:B} + 3\beta_{AB_3} \quad (\mathrm{A.10'})$$

$$L^{\mathrm{4SL}}_{\mathrm{A,B:A,B:*:*}} = w_{\mathrm{A:B}} \quad (\mathrm{A.11'})$$

Above-mentioned discussion can be extended to ternary system, for example;

$$G^{4SL}_{A_2BC} = 2w_{AB} + 2w_{AC} + w_{BC} + \gamma_{A_2BC} \quad (A.12)$$

$$G^{4SL}_{AB_2C} = 2w_{AB} + 2w_{BC} + w_{AC} + \gamma_{AB_2C} \quad (A.13)$$

$$G^{4SL}_{ABC_2} = 2w_{AC} + 2w_{BC} + w_{AB} + \gamma_{ABC_2} \quad (A.14)$$

$$L^{4SL}_{A,B:A,C:*:*} = 0.5(w_{A:B} + w_{A:C} - w_{B:C} - w_{A:A}) \quad (A.15)$$

$$L^{4SL}_{A,B:B,C:*:*} = 0.5(w_{A:B} + w_{B:C} - w_{A:C} - w_{B:B}) \quad (A.16)$$

$$L^{4SL}_{A,C:B,C:*:*} = 0.5(w_{A:C} + w_{B:C} - w_{A:B} - w_{C:C}) \quad (A.17)$$

where $\gamma_{A_2BC}$ is correction factor.

In the present study, the formation enthalpy values of the stoichiometric compounds, for example in the A-B binary system; $G^{4\mathrm{SL}}_{A_3B}$, $G^{4\mathrm{SL}}_{A_2B_2}$ and $G^{4\mathrm{SL}}_{AB_3}$, are calculated by first principles calculations, which are applied to estimate *w*, *α*, *β* and *γ* parameters in eqs. (A.8’)-(A.17).

| Structure | Lattice Distances | $d/a$ (ideal ratio :$\sqrt{2/3}$~0.8165) | $\Delta(d/a)$ |
|---|---|---|---|
| 2H | $a = 3.188$Å | | |
| | $d = 2.595$Å | 0.8141 | -0.3% |
| 3R | $a = 3.196$Å | | |
| | $d = 2.604$Å | 0.8147 | -0.2% |
| 14H | $a = 3.195$Å | | |
| | $d_1 = 2.599$Å | 0.8136 | -0.4% |
| | $d_2 = 2.597$Å | 0.8128 | -0.5% |
| | $d_3 = 2.595$Å | 0.8122 | -0.5% |
| | $d_4 = 2.597$Å | 0.8129 | -0.4% |

**Table 1**

The interatomic and interlayer distances, and their aspect ratios of the 2H, 3R and 14H structures in Fig. 1.

| Phase and model | Thermodynamic parameters, J/mol |
|---|---|
| $fcc$: (Mg, Y, Zn) (Va) | ${}^{0}L^{\mathrm{fcc}}_{\mathrm{Mg,Y:Va}} = -19148.97108 + 4.3713048848T$<br>${}^{1}L^{\mathrm{fcc}}_{\mathrm{Mg,Y:Va}} = +7520.16709$<br>${}^{2}L^{\mathrm{fcc}}_{\mathrm{Mg,Y:Va}} = +13475.15536$<br>${}^{0}L^{\mathrm{fcc}}_{\mathrm{Mg,Zn:Va}} = -16219.3811 + 1.63049887T$<br>${}^{1}L^{\mathrm{fcc}}_{\mathrm{Mg,Zn:Va}} = +3914.41115 - 0.5391724715T$<br>${}^{2}L^{\mathrm{fcc}}_{\mathrm{Mg,Zn:Va}} = +4208.65809 - 0.5758194T$<br>${}^{0}L^{\mathrm{fcc}}_{\mathrm{Y,Zn:Va}} = -143756.7032 + 9.88T$<br>${}^{1}L^{\mathrm{fcc}}_{\mathrm{Y,Zn:Va}} = +36802.82107$<br>${}^{2}L^{\mathrm{fcc}}_{\mathrm{Y,Zn:Va}} = +23970.19655$<br>${}^{0}L^{\mathrm{fcc}}_{\mathrm{Mg,Y,Zn:Va}} = +214939 - 507T$<br>${}^{1}L^{\mathrm{fcc}}_{\mathrm{Mg,Y,Zn:Va}} = +290411 - 582T$<br>${}^{2}L^{\mathrm{fcc}}_{\mathrm{Mg,Y,Zn:Va}} = -6593 + 4.6T$ |
| $hcp$: (Mg, Y, Zn) (Va) | ${}^{0}L^{\mathrm{hcp}}_{\mathrm{Mg,Y:Va}} = -10423.5191 - 3.101141295T$<br>${}^{1}L^{\mathrm{hcp}}_{\mathrm{Mg,Y:Va}} = -8707.63339$<br>${}^{2}L^{\mathrm{hcp}}_{\mathrm{Mg,Y:Va}} = -754.279598$<br>${}^{0}L^{\mathrm{hcp}}_{\mathrm{Mg,Zn:Va}} = -7748.038178$<br>${}^{1}L^{\mathrm{hcp}}_{\mathrm{Mg,Zn:Va}} = +9085.6193$<br>${}^{2}L^{\mathrm{hcp}}_{\mathrm{Mg,Zn:Va}} = +327.6354$<br>${}^{0}L^{\mathrm{hcp}}_{\mathrm{Y,Zn:Va}} = -122504.6251$<br>${}^{1}L^{\mathrm{hcp}}_{\mathrm{Y,Zn:Va}} = +72194.124$<br>${}^{2}L^{\mathrm{hcp}}_{\mathrm{Y,Zn:Va}} = +9108.042$<br>${}^{0}L^{\mathrm{hcp}}_{\mathrm{Mg,Y,Zn:Va}} = +221308$<br>${}^{1}L^{\mathrm{hcp}}_{\mathrm{Mg,Y,Zn:Va}} = -163679$<br>${}^{2}L^{\mathrm{hcp}}_{\mathrm{Mg,Y,Zn:Va}} = -82202$ |

**Table 2**

Thermodynamic parameters of *hcp* and *fcc* disordered solution phases evaluated in the present work. First principles calculations in CE-CVM were performed using VASP under the conditions of a plane-wave energy cutoff of 400 eV with 2000 *k*-points per reciprocal atoms.

| | | | | | | |
|---|---|---|---|---|---|---|
| $L1_2$-$A_3B$ | $Mg_3Y$ | -7.71 kJ/mol | $Mg_3Zn$ | -2.41 kJ/mol | $Y_3Zn$ | -2.74 kJ/mol |
| | $MgY_3$ | +1.75 kJ/mol | $MgZn_3$ | -4.31 kJ/mol | $YZn_3$ | -21.93 kJ/mol |
| $L1_0$-AB | MgY | -5.02 kJ/mol | MgZn | -5.37 kJ/mol | YZn | -20.83 kJ/mol |
| *F'*-$A_2BC$ | $Mg_2YZn$ | -12.3 kJ/mol | $MgY_2Zn$ | -11.4 kJ/mol | $MgYZn_2$ | -18.0 kJ/mol |

**Table 3**
Calculated formation enthalpies of the various ordered-*fcc* structures at 0 K in the Mg-Y, Mg-Zn and Zn-Y binary and Mg-Zn-Y ternary systems.

| | | | | | |
|---|---|---|---|---|---|
| $w_{\mathrm{Mg:Y}}$ | $-1255$ J/mol | $w_{\mathrm{Mg:Zn}}$ | $-1342$ J/mol | $w_{\mathrm{Y:Zn}}$ | $-5208$ J/mol |
| $\alpha_{\mathrm{Mg_3Y}}$ | $-1315$ J/mol | $\alpha_{\mathrm{Mg_3Zn}}$ | $+538$ J/mol | $\alpha_{\mathrm{Y_3Zn}}$ | $+4295$ J/mol |
| $\beta_{\mathrm{MgY_3}}$ | $+1803$ J/mol | $\beta_{\mathrm{MgZn_3}}$ | $-94$J/mol | $\beta_{\mathrm{YZn_3}}$ | $-2100$ J/mol |
| $\gamma_{\mathrm{Mg_2YZn}}$ | $-1903$ J/mol | $\gamma_{\mathrm{MgY_2Zn}}$ | $+2903$ J/mol | $\gamma_{\mathrm{MgYZn_2}}$ | $-3639$J/mol |

**Table 4**

Calculated values of bond energies and correction terms. Details of the parameters $\mathrm{w}, \alpha, \beta$ and $\gamma$ are explained in Appendix.

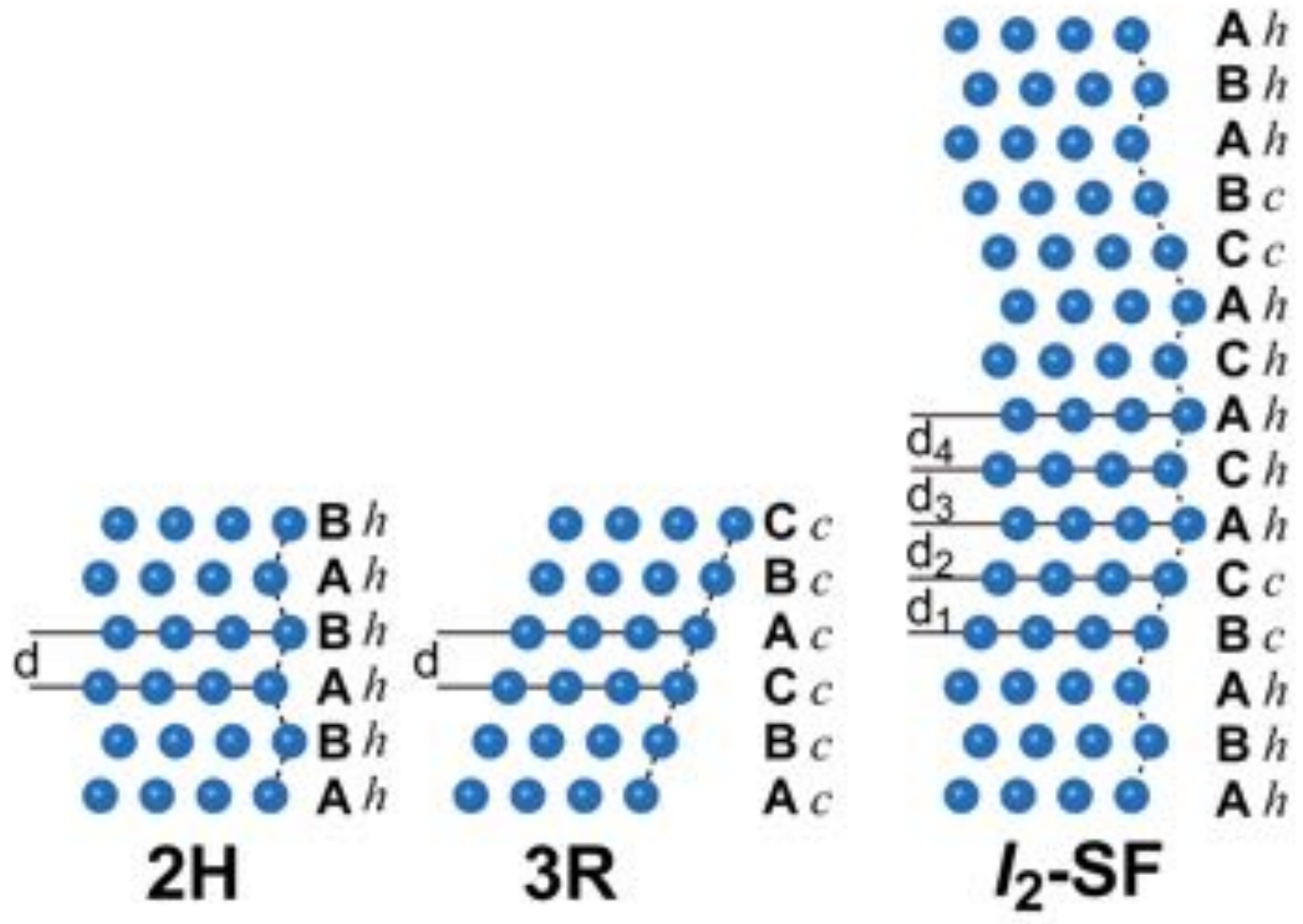

**Figure 1**

Pure Mg stacking structures of 2H (*hcp*), 3R and 14H-type composed of $I_2$-SFs. For the structure optimization, VASP within the framework of the density functional theory (DFT) [30], based on the generalized gradient approximation (GGA) [31] and the ultra-soft scalar relativistic pseudo-potential with plane-wave energy cutoff of 500 eV were employed. K-point meshes of $39 \times 39 \times 21$, $41 \times 41 \times 41$ and $39 \times 39 \times 3$ were chosen for 2H, 3R and 14H, respectively, and a Methfessel-Paxton smearing method [51] with a width of 0.2 eV was employed.

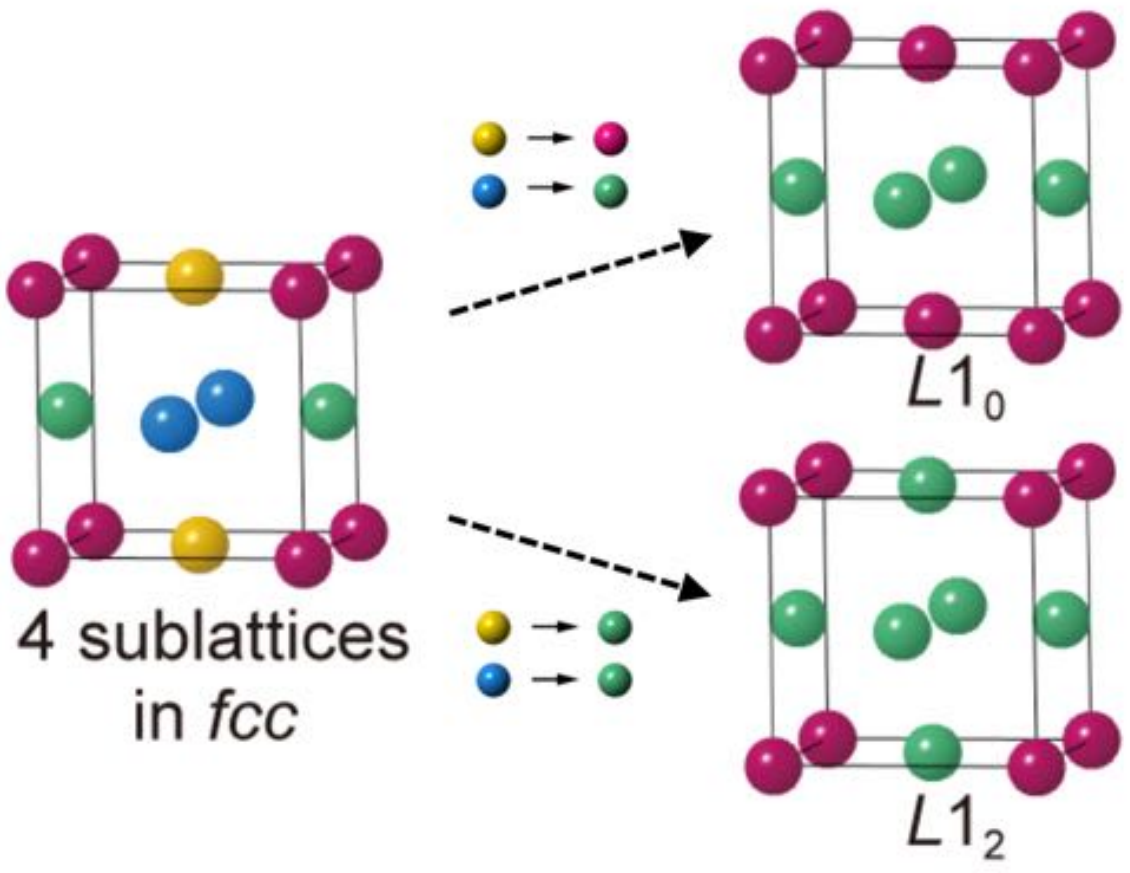

**Figure 2**

Schematic illustration of a four-sublattice model, by which Gibbs energy of both the $L1_0$ and the $L1_2$ structures can be described.

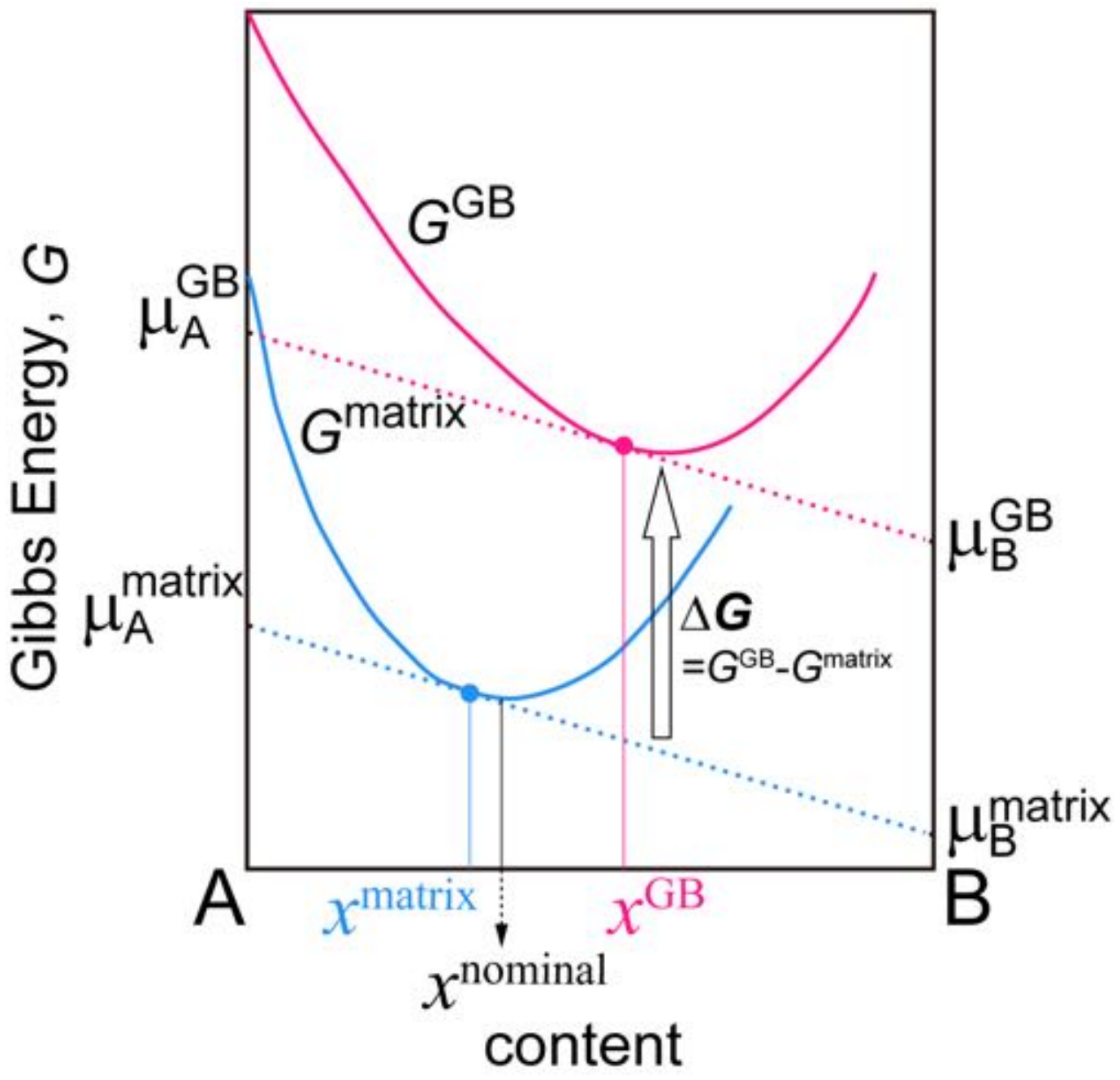

**Figure 3**

Schematic illustration of the Hillert's parallel tangent law for the evaluation of grain boundary (GB) segregations in the A-B binary system. $\mu_i^{\theta}$ represents the chemical potential of element *i* in the phase θ, and $x^{\theta}$ means the composition of the phase θ.

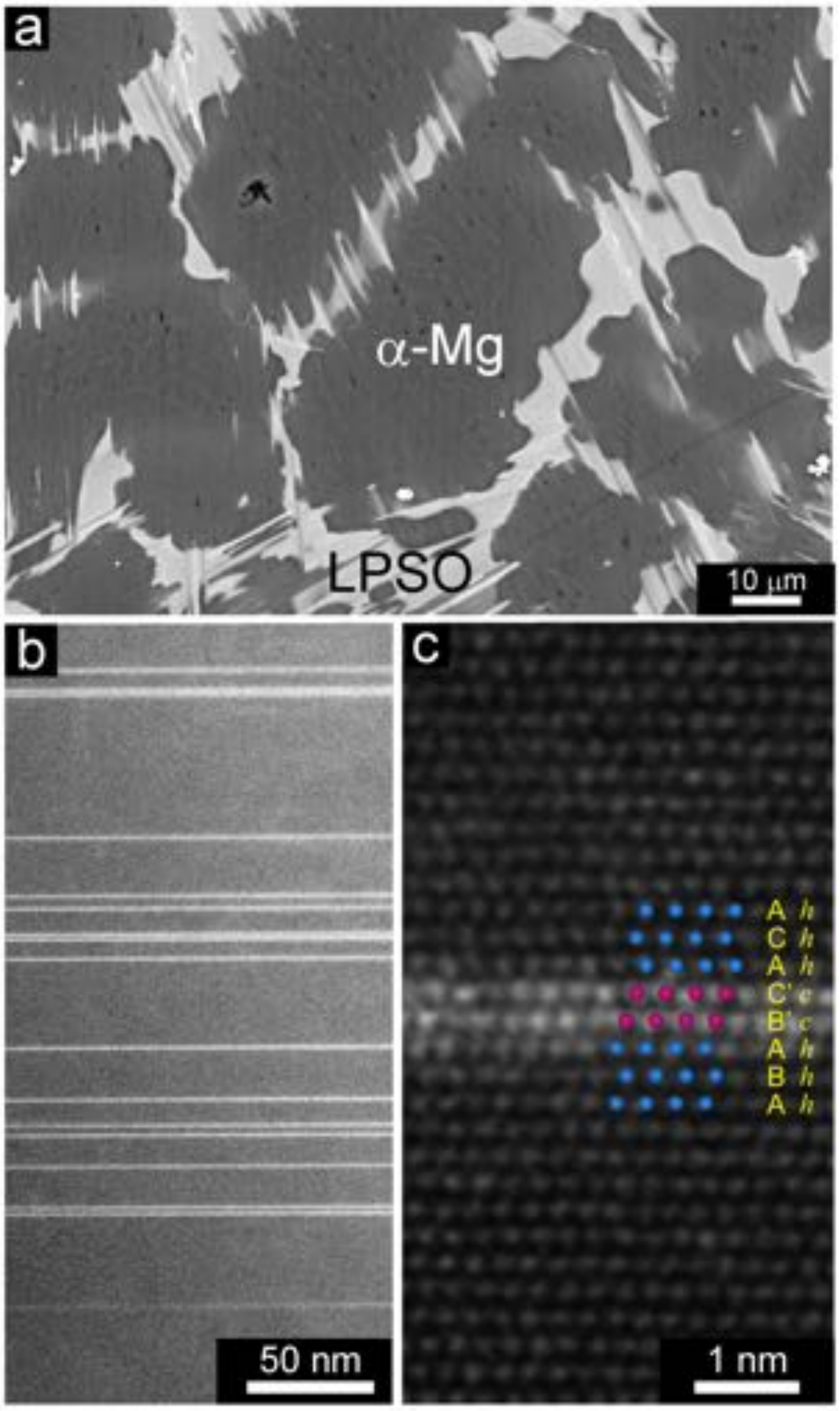

**Figure 4**

(a) SEM and (b, c) HAADF-STEM images of the $Mg_{97}Zn_1Y_2$ alloy annealed at 523 K for a week. Bright lines in (b) represent the *fcc*-SESF at which the Zn/Y atoms segregate. In the atomic resolution HAADF image (c), it is clearly seen that the significant Zn/Y enrichment occurs at the local *fcc* environment.

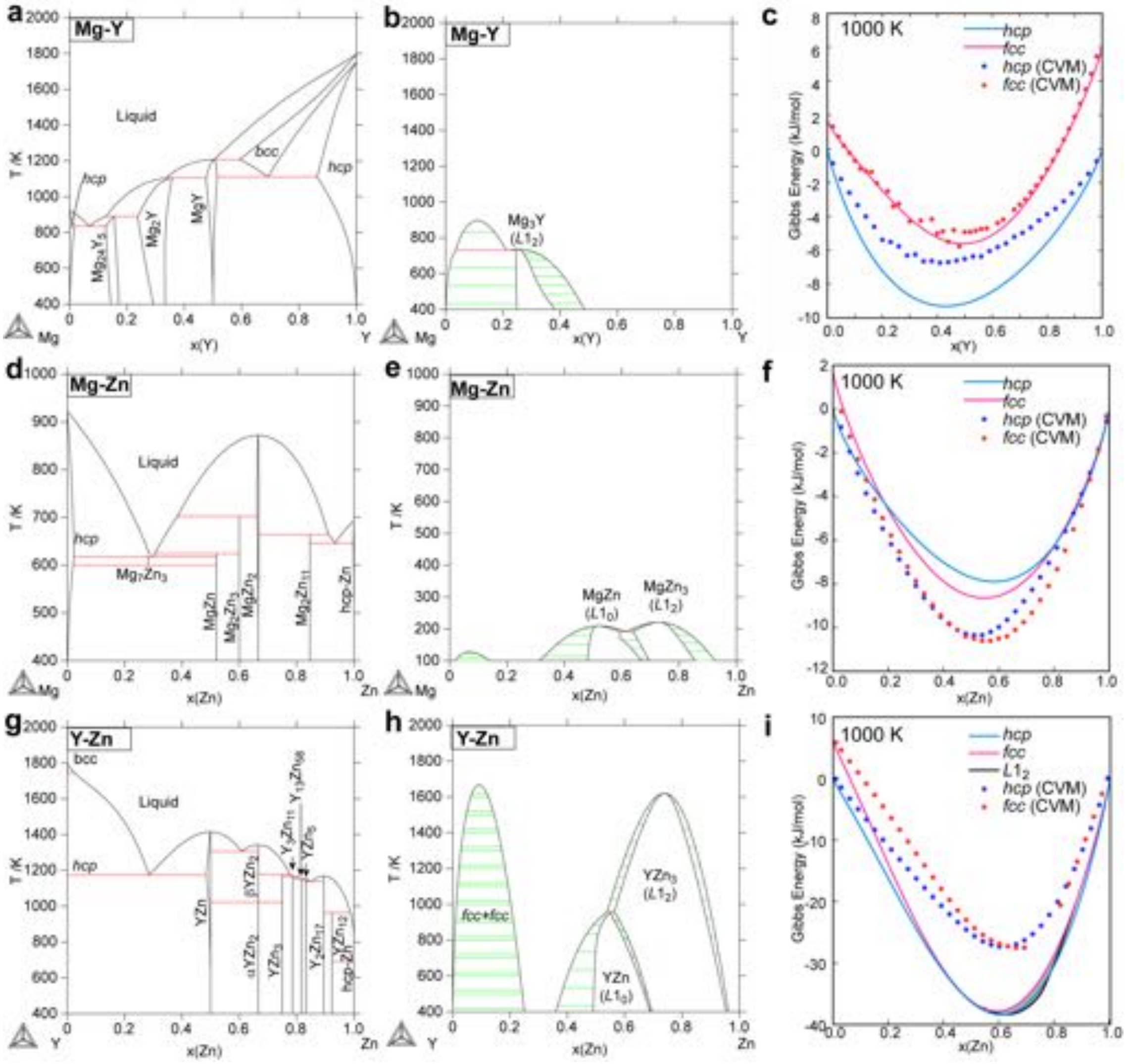

**Figure 5**

Calculated (a-c) Mg-Y, (d-f) Mg-Zn and (g-i) Zn-Y binary phase diagrams and Gibbs energy curves. (a, d, g) stable phase diagrams, (b, e, h) metastable phase diagrams with the *fcc*, $L1_2$ and $L1_0$ phases and (c, f, i) Gibbs energy curves of the *hcp* and *fcc* (solid lines) compared with the calculated values by CE-CVM (dotted lines).

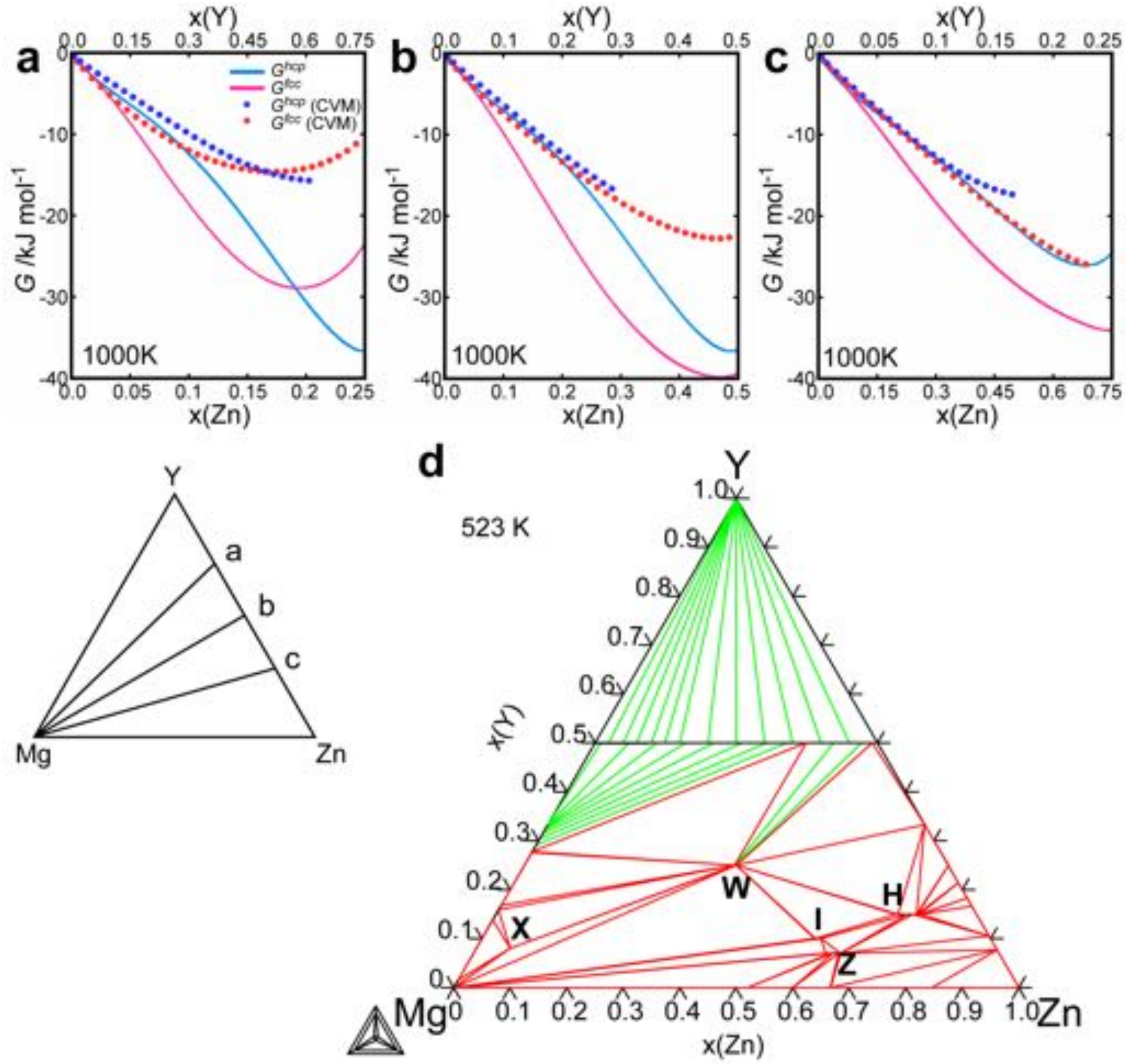

**Figure 6**

(a-c) Calculated cross-sectional Gibbs energy curves of the solid-solution *hcp* and *fcc* phases across a, b and c in the schematic Mg-Zn-Y ternary diagram at the lower-left. CE-CVM results are shown together by dotted lines. (d) Stable phase diagram of the Mg-Zn-Y ternary system calculated using the thermodynamic parameters in the present work. Ternary compounds X, W, I, Z and H represent $Mg_{12}ZnY$, $Mg_3Zn_3Y_2$, $Mg_3Zn_6Y$, $Mg_{28}Zn_{65}Y_7$ and $Mg_{15}Zn_{70}Y_{15}$, respectively (details are in reference [23]).

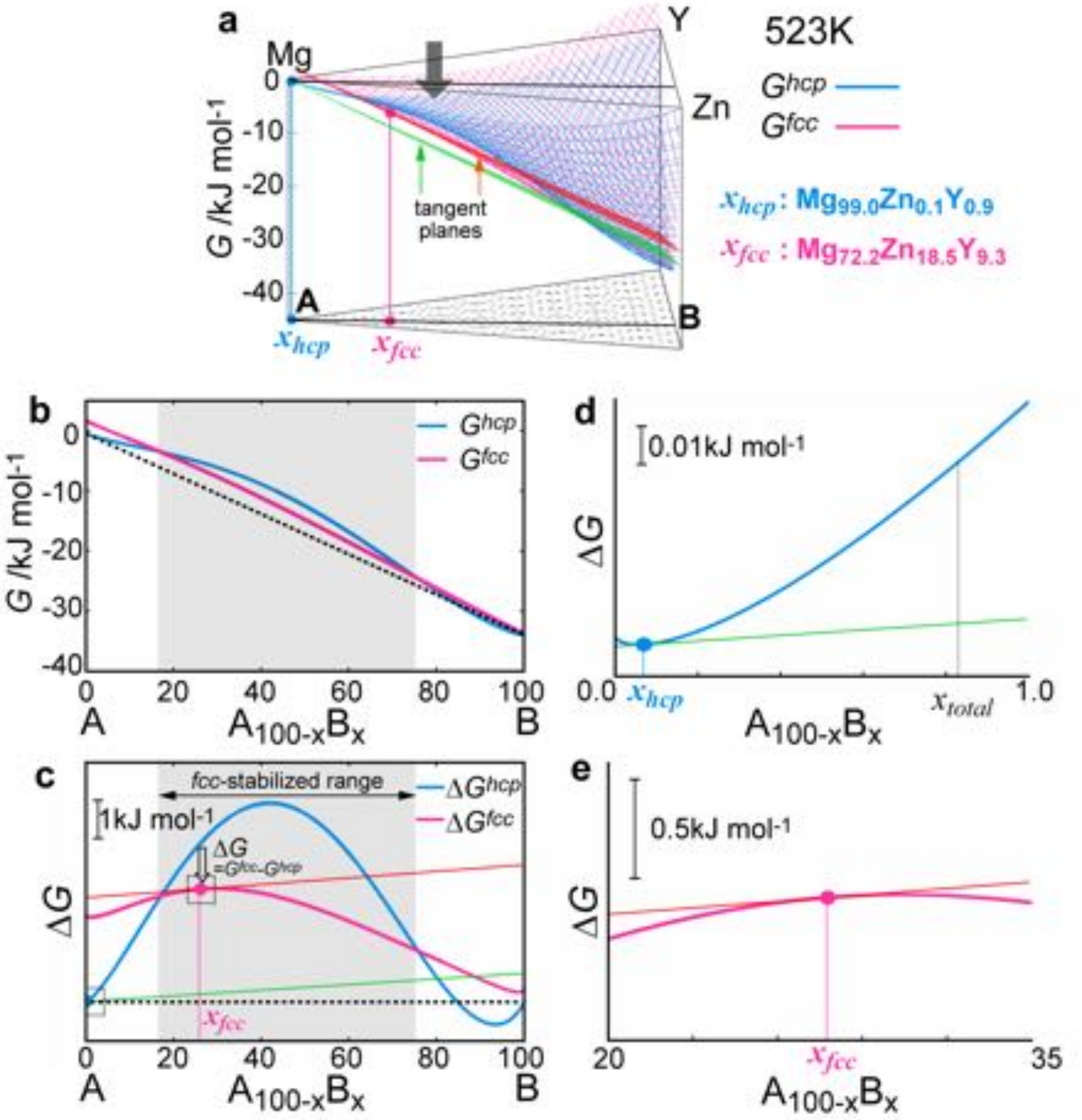

**Figure 7**

(a) Calculated Gibbs energy surfaces of *hcp* (blue) and *fcc* (light-red) phases, shown together with the parallel tangent planes (green and red, respectively). The compositions determined at the tangent points for the *hcp* and *fcc* phases are indicated as $x_{hcp}$ and $x_{fcc}$, respectively. (b) Cross sectional Gibbs energy curves of (a) across the tangent points, and (c) the redrawn *G* curves by taking the dot-line in (b) as flat (i.e., $G^{hcp}$ to be equivalent at the both sides of A and B). The compositions at the both ends are A: $Mg_{99.0}Y_{1.0}$ and B: $Mg_{0.1}Zn_{70.5}Y_{29.4}$. (d), (e) enlargement around the tangent point for the *hcp* and the *fcc* phases in (c), respectively.

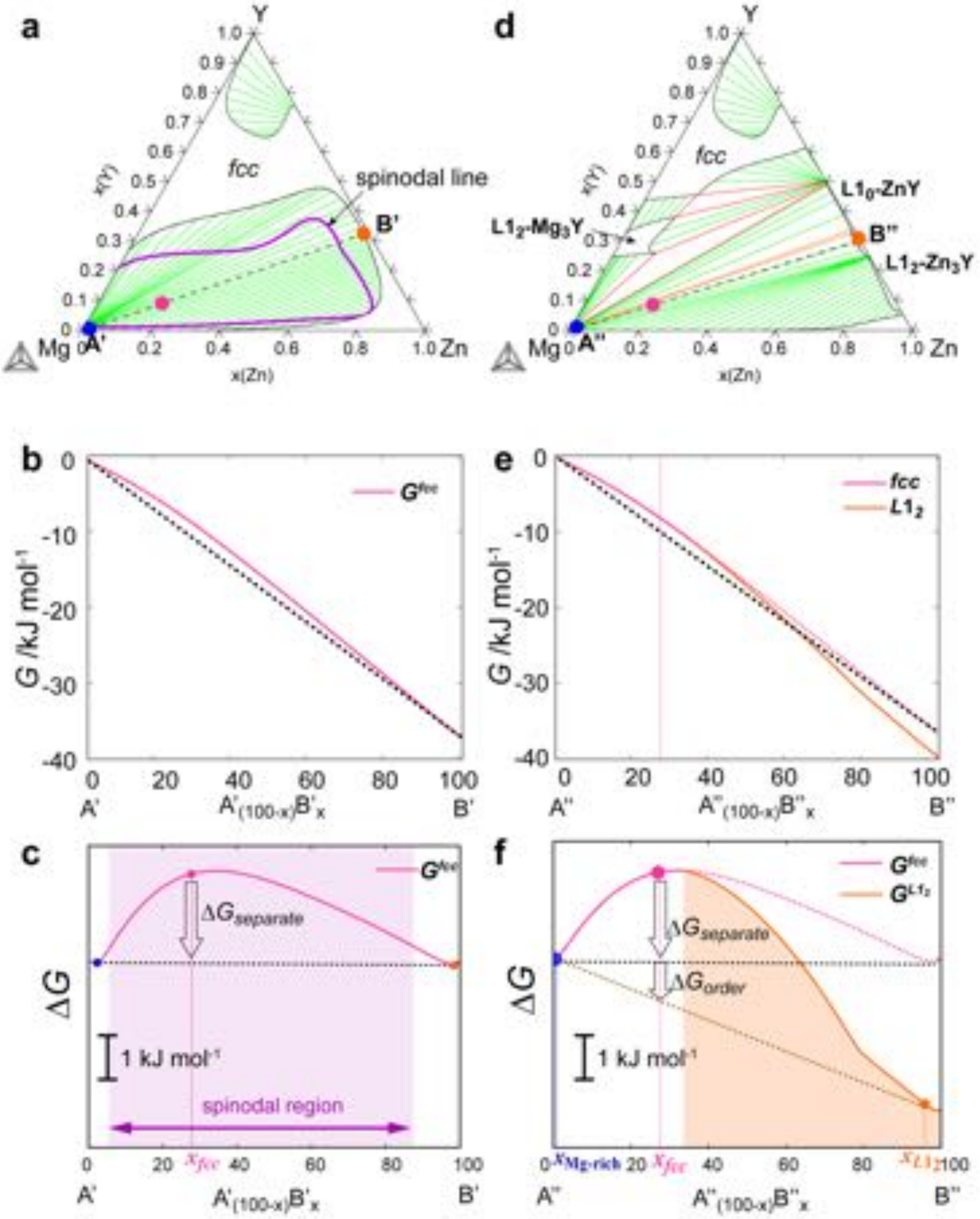

**Figure 8**

Isothermal diagrams and Gibbs energy curves of the meta-stable *fcc* Mg-Zn-Y phases calculated at 523K; (a-c) with the disorder *fcc* phase, (d-f) with the disorder *fcc*, ordered $L1_0$ and $L1_2$ phases. (a), (d) Isothermal section of the meta-stable Mg-Zn-Y ternary diagrams. (b), (e) Cross sectional Gibbs energy curves across A'-B' in (a) and A"-B" in (b), respectively. (c), (f) Redrawn Gibbs energy curves by taking the dot-line in (b) and (e) as flat, along the same manner used for the Figs. 7 (b) and (c). Original composition segregated at the *fcc* phase ($x_{fcc}$, light-red) spontaneously separate into two *fcc* phases by spinodal decomposition, Mg-rich (blue) and Zn/Y-rich (red) shown in (c), and the latter further undergoes disorder-order transformation to form the $L1_2$-ordered phase (tangerine) shown in (f).

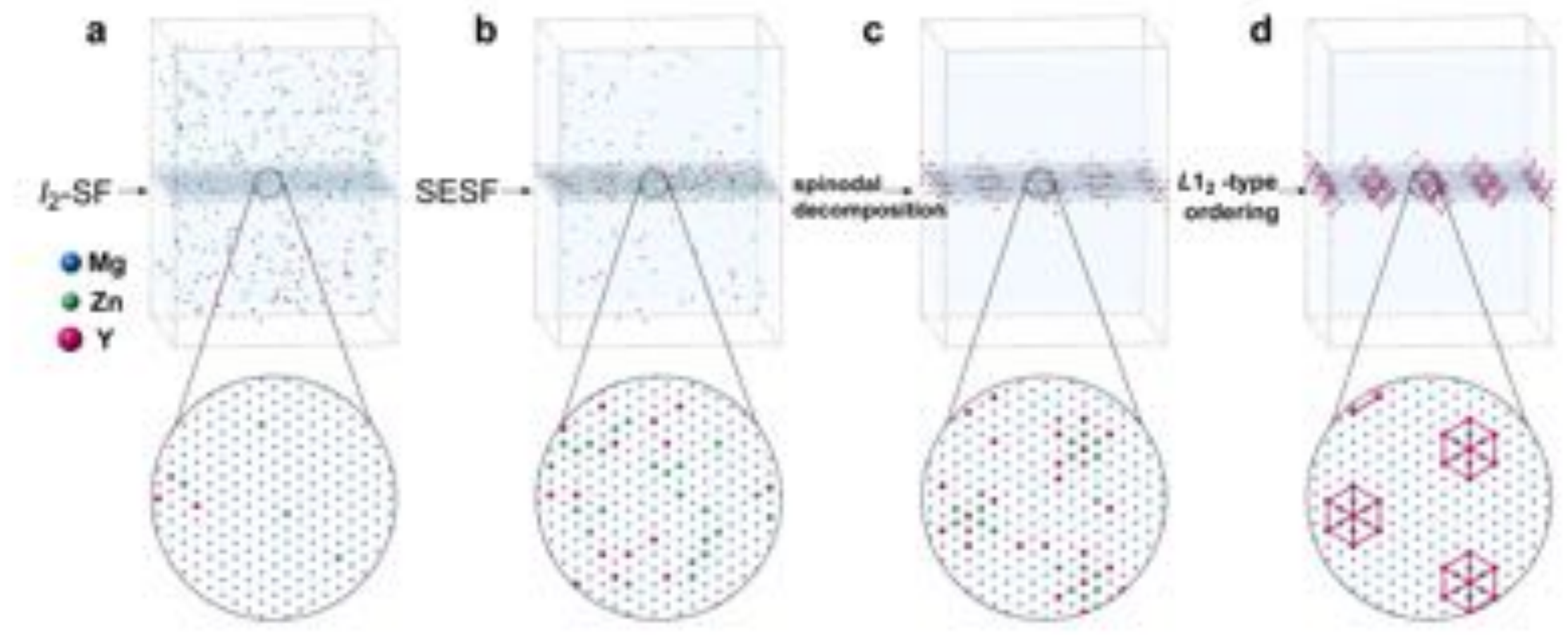

**Figure 9**

Phenomenological sequences of the solute partitioning behaviors around the stacking faults in dilute Mg-Zn-Y alloys. (a) $I_2$-SFs are introduced, where (b) the solute Zn/Y atoms are segregated to form the *fcc*-SESF, followed by (c) spinodal decomposition between Mg-rich and Zn/Y-rich regions, (d) the latter of which undergoes disorder-order transformation to form the $L1_2$-type order phase. Each of these events takes place to reduce the total Gibbs energy of the system.

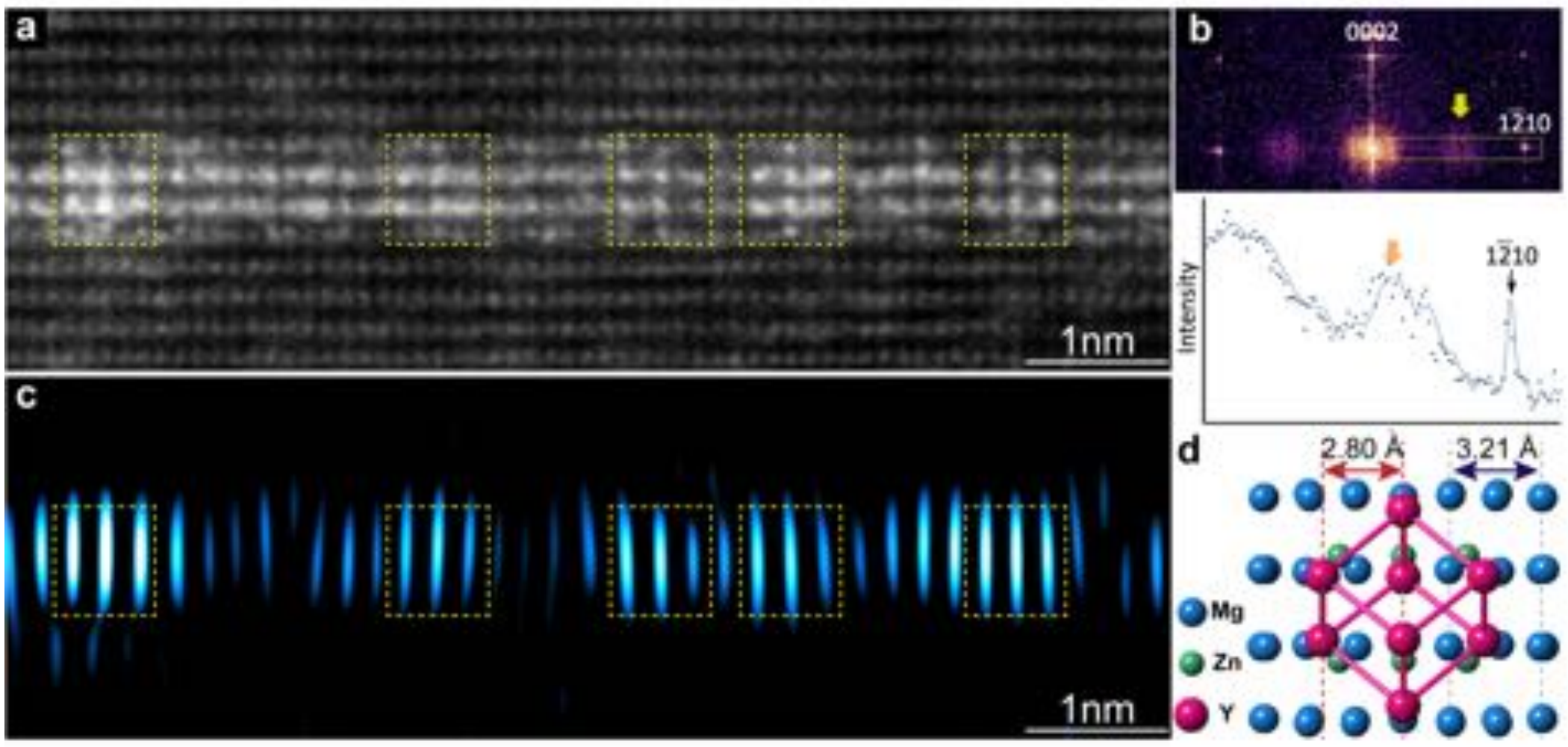

**Figure 10**

(a) Atomic-resolution HAADF-STEM image of the SESF taken along the $[10\bar{1}0]_{hcp}$ direction. (b) FFT pattern obtained from (a), and the intensity profile along the dotted-box region is shown below. (c) Inverse FFT image reproduced using the diffuse scattering indicated by an bold arrow in (b), whose peak is at an approximately 2.8Å correlation-length. (d) $L1_2$-type short-range order $Zn_6Y_8$ cluster embedded in the *fcc*-SESF layers; the Zn/Y configurations are significantly relaxed after energetic optimizations [7].